# How does the Monte Carlo method work?

Oleg Yavoruk

**Abstract**
The paper describes the practical work for students visually clarifying the mechanism of the Monte Carlo method applying to approximating the value of Pi. Considering a traditional quadrant (circular sector) inscribed in a square, here we demonstrate the original algorithm for generating random points on the paper: you should arbitrarily tear up a paper blank to small pieces (the first experiment). By the similar way the second experiment (with a preliminary staining procedure by bright colors) can be used to prove the quadratic dependence of the area of a circle on its radius. Manipulations with tearing up a paper as a random sampling algorithm can be applied for solving other teaching problems in physics.

## Introduction

Monte Carlo is one of Europe's leading tourist resorts, an administrative area of the Principality of Monaco where the famous Monte Carlo Casino is located.

A group of numerical methods using random processes has the same name. This term was suggested by Nicholas Metropolis directly bearing in mind his relative who was very passionate about gambling: the most reliable random number generator is roulette exactly, a kind of gambling.

The first description of this method appeared in 1949 [1]. Later computers significantly expanded the range of problems that this method solved effectively. Remarkably, it was used to create a thermonuclear weapon in the 1950s. And such a famous and verified numerical method is astoundingly not mentioned in the traditional school and university physics courses.

Its fields of application are a technique for numerical integration, solving of algebraic linear and nonlinear equations, differential and integral equations, modeling of natural processes. It is used in mechanics, aerodynamics, molecular physics, quantum physics, solid state physics, plasma physics, and astrophysics. Monte Carlo algorithms are a classical tool to demonstrate the time evolution of some processes occurring in nature, to analyze physical systems [2], to calculate moments of inertia [3], to simulate a rainbow [4], to show statistical fluctuations of the radioactive decay [5]. A Monte Carlo method is used to illustrate some of the principles of statistical mechanics: the concepts of ensembles, statistical averages, fluctuations [6], diffusion [7].

This method has shown reliability beyond physics: in mathematical finance, the calculation of risks in business, engineering problems, computational biology, computer graphics, applied statistics, artificial intelligence and even modern weather forecasting. Sometimes it is the unique method for solving a problem in a reasonable time [8, 9, 10, 11]. Versions of this method have long been used in sociology, political science, logic, linguistics, psychology and pedagogy [12, 13, 14].

In our lab we should figure out the mechanism of this method on a very simple example. The first experiment is devoted to calculating the number $\pi$ with random tests. In the second experiment we are going to prove the formula $A=\pi \cdot R^2$ (quadratic dependence of the area of a circle on the radius). Experiments can be carried out in any order, as well as independently from each other.

## What is Pi?

The number π (Pi) is a mathematical constant, the ratio of the circumference of a circle to its diameter. Pi is infinite decimal.

The number π is important not only to mathematicians; it plays a big role in various branches of physics (the body rotation, oscillations and waves, physics of light, atomic and nuclear physics, quantum mechanics, elementary particle physics, etc.) [15, 16].

In practice, we only need to know a few first decimal digits, but sometimes we need more accuracy [17]. However, for most cosmological calculations, 39-40 digits are enough, because it is the accuracy needed to calculate the circumference of the Observable Universe with an accuracy of an atom. Value π = 3.141592653589793, rounded to 15 characters, NASA uses for the planetary navigation problems [18].

There are various ways to calculate the number π: geometrical, numerical, analytical, experimental, computer [19]. And there are algorithms based on the Monte Carlo method, related to: throwing a needle on a lined paper, so called the Buffon Method [20, 21], shooting with a rifle shot [22], throwing darts [23, 24], etc.

But we are going to perform a version based on a comparison of the areas of a circle and a square, exploiting another practical algorithm for obtaining the results of random tests: manipulations with tearing up a paper.

## Experiment 1. Calculation of Pi

The general technique of Monte Carlo simulations described in several papers [25, 26, 27, 28].

Suppose we have a flat figure (it is a circle in our case) with the area $A_0$ which we need to find. Restrict it to another figure with the area $A_1$ (in our case it's a square). Thus we draw a circle inscribed in a square (Figure 1a).

The area of the square $A_1 = d^2$, $d$ is a side of the square. Diameter of our circle equals to the side of a square. The area $A_0$ of a circle is $\pi \cdot R^2$. Diameter $d = 2 \cdot R$. And then $A_0 = \pi \cdot d^2/4$. The ratio of the areas of a circle and a square is:

$$\frac{A_0}{A_1} = \frac{\pi \cdot \frac{d^2}{4}}{d^2} = \frac{\pi d^2}{4d^2} = \frac{\pi}{4}$$

The essence of the Monte Carlo method is very simple. If we allocate points randomly within a square (Figure 1b), the ratio of the areas of a circle and a square is equal to the ratio of the number of points $N_0$ (that fall into *a circle*) and the total number of points $N_1$:

$$\frac{A_0}{A_1} = \frac{N_0}{N_1}$$

The larger the area, the more points it gets.

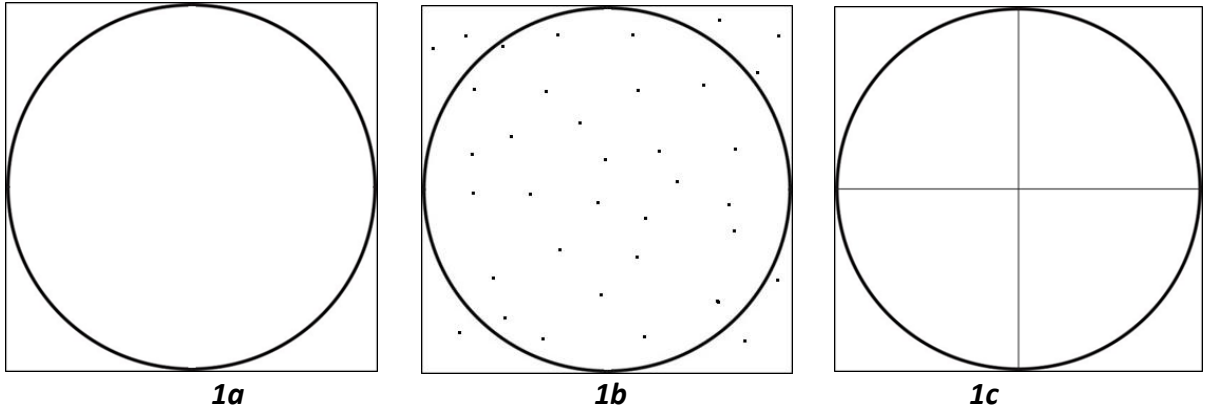

*1a*  *1b*  *1c*

*Figure 1 (a, b, c)*

In addition, the ratio of the areas will not change if the object is cut into four equal parts (Figure 1c). We can divide our drawing into four parts (quadrants) as shown in the figure. And now we have the opportunity to conduct four tests using only one blank.

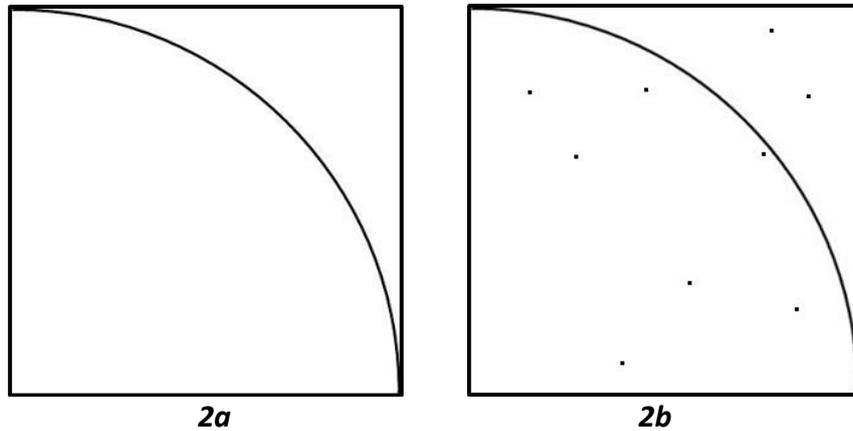

*2a*  *2b*

*Figure 2 (a, b)*

This is a wonderful gift: we can conduct four tests (Figures 2a and 2b) with one drawing. Of course, we can do more tests. The Appendix 1 (at the end of the paper) presents the blank with five tests.

Based on the above reasoning about the number of randomly distributed points inside the figures, we see that with a large number of tests the number π is equal

$$\pi = 4 \cdot \frac{N_0}{N_1},$$

$N_0$ - the number of points inside the circle, $N_1$ - the number of points inside the square (including the circle).

The next question is: how to randomly arrange the points inside a square containing an inscribed circle? Of course modern computers can help us to generate random coordinates.

But here are opportunities for fans of natural experiments: scattering rice grains on the drawn square; random distributions of small coins or buttons;

raindrops on a paper sheet; dart throwing; and even shooting at a target (a square with an inscribed circle) with a shotgun. However, in this lab we are going to use a distinctly visual and pretty simple method. Not a lot of people know about this way.

The experimenter paints two areas with radically different colors (the simplest examples are white and black), and then randomly *breaks (tears) it up* into small pieces.

Figure 3a presents the results of this experiment. Now we have two groups of pieces: almost white (from the inside of our circle) and almost black (from the outside of the circle). The total number of pieces: $N_1$, white: $N_0$, black: $(N_1 - N_0)$.

Here we are going to carry out four tests. Therefore, we repeat it three times more (Figures 3b, 3c, 3d).

The numerous small pieces of paper are assembled as a puzzle only for illustrative purposes and exclusively for this paper (Figure 3). Picking up puzzles is an exciting activity, but it is not mandatory here: it is enough just to count the pieces of paper.

If the black and white areas on one piece are approximately equal, it does not matter, should we consider this piece white or black: our experimenter selects one of them. If there are a few such pieces, some of them can be attributed to white, some to black (equal parts), and the last (odd) one to any group. If you are unsure about a last piece, you can just break it in half: it does not matter for a large number of pieces after all.

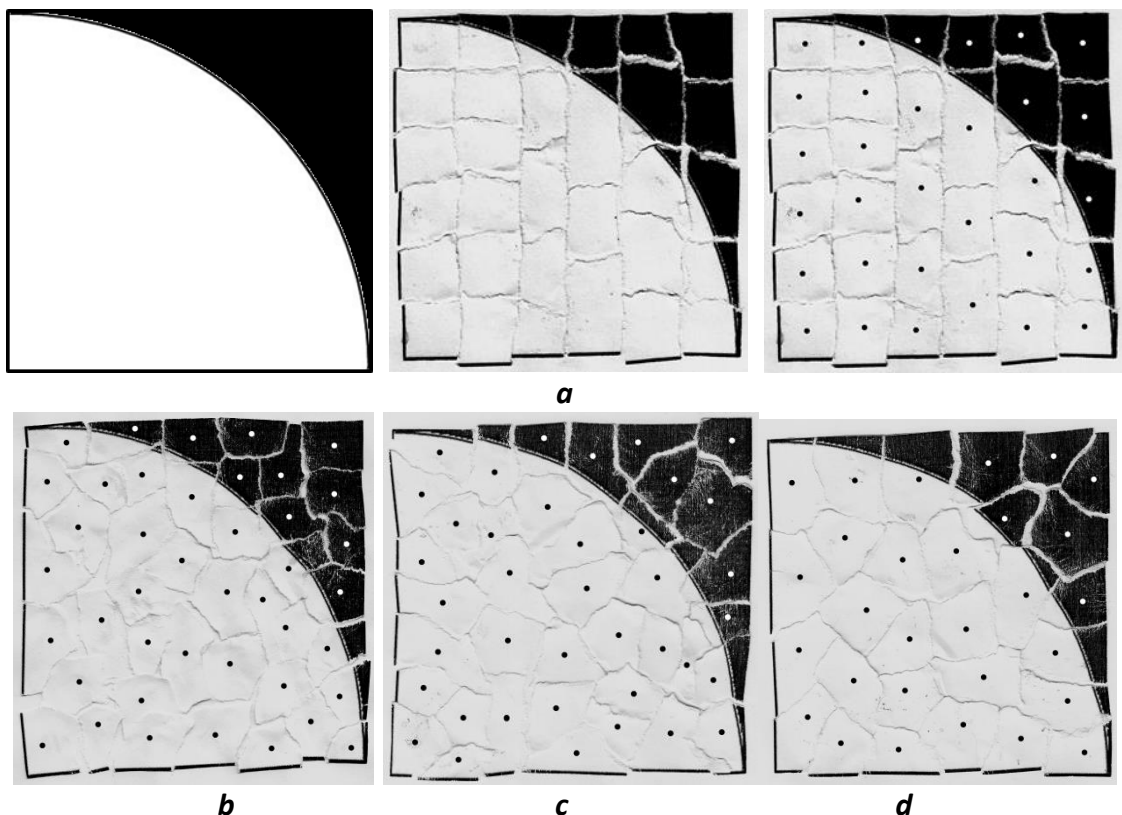

*a*

*b*         *c*         *d*

**Figure 3 (a, b, c, d)**

Then we fill in the table with the results of all four tests; calculate the arithmetic average value <π>, standard interval *s* and absolute error *Δπ*.

**Table 1. The results of the first experiment**

| k | $N_1$ | $N_0$ | $\pi_k = 4 \cdot \dfrac{N_0}{N_1}$ | $\|<\pi> - \pi_k\|$ | $(<\pi> - \pi_k)^2$ |
|---|---|---|---|---|---|
| 1 | 24+7 = 31 | 24 | 3.10 | 0.02 | 0.0004 |
| 2 | 32+10 = 42 | 32 | 3.04 | 0.08 | 0.0064 |
| 3 | 31+8 = 39 | 31 | 3.18 | 0.06 | 0.0036 |
| 4 | 23+6 = 29 | 23 | 3.17 | 0.05 | 0.0025 |
| $k_{max}$ = 4 | | | <π> = 3.12 | | $\sum_{k=1}^{N}(\langle\pi\rangle - \pi_k)^2 =$ 0.0129 |

There are various ways to estimate measurement uncertainties. For example, this can be done on the basis of a given statistical confidence and the number of tests [29, 30]. We are going to evaluate the measurement error *Δπ* as follows:

$$\Delta\pi = t_{p,N} \cdot S,$$

Student's *t*-factor $t_{p,N}$ (for $k_{max}$ = 4 and *p* = 90%) equals *2.35* [31].
The experimental standard deviation *S* in our case [29]:

$$S = \sqrt{\dfrac{\sum_{k}(<\pi> - \pi_k)^2}{N(N-1)}} \approx 0.03$$

And finally the absolute uncertainty *Δπ* ≈ *0.08* (the relative uncertainty *ε* is a little more 2%). The calculated confidence interval:

*π = 3.12 ± 0.08 (with probability p = 90%).*

Of course, here we already know in advance the result of the experiment (*π*) which should be obtained in principle with great precision. This is a rare pedagogical luck: very few researchers possess a previously known answer. Perhaps next time we'll not be so lucky. Taking this opportunity here we should clearly show the cases where our result falls inside the confidence interval, as well as outside the one. The procedure of comparing the experimental $\pi_э$ and theoretical results *π* has a great educational value.

We can mark experimental results on the linear axis. Let's consider following examples.

1. $\pi_э = 3.18 \pm 0.06$  $p = 90\%$  ($\pi$ is inside the confidence interval)

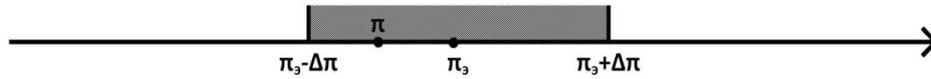

2. $\pi_э = 3.52 \pm 0.09$  $p = 90\%$  ($\pi$ is outside the confidence interval)

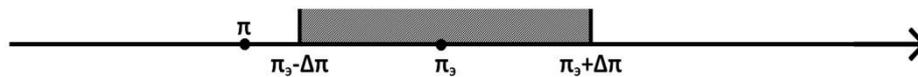

3. $\pi_э = 2.52 \pm 0.04$  $p = 90\%$  ($\pi$ is again outside the confidence interval)

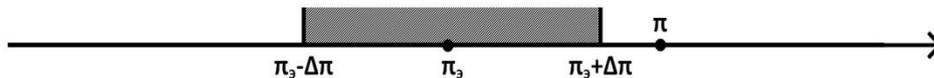

We can hope that with an increase in the number of tests ($N > 4$), the accuracy of calculating the number $\pi$ will increase too. The confidence interval can be changed by increasing or decreasing the value of the probability $p$.

Similar explanations and clarifications provide amazing opportunities for informative playing with confidence intervals, probabilities, uncertainties, the number of data values, and ensuing discussion on the relation between them.

**Experiment 2. The proof of a quadratic dependence of the area on the radius:**
$$A = \pi \cdot R^2$$

The Monte Carlo method can also be applied for the confirmation or refutation of different theoretical conjectures. We are going to illustrate this studying the dependence of the area of a circle on its radius. Our theory asserts the existence of a quadratic dependence $A(R)$: a parabola with a vertex at the origin.

We draw five concentric circles (Figure 4a). The fifth (outer) circle is inscribed in a square like in the previous experiment.

A person of any age enjoys looking at a brightly colored drawing. Especially if the experimenter has the ability to choose different colors (Figure 4b). However, you should not forget to paint over the small rectangles in Table 2 exactly like in Figure 4b (by the same colors), since after transforming (tearing up) of our beautiful picture into small pieces, this will be a little more difficult (Figure 4c).

As in the previous experiment, the puzzle has been solved only for this paper, and exclusively for a spectacular illustration. This time-consuming (but exciting) activity is absolutely optional here. You just need to carefully count the resulting pieces of paper. Each one symbolizes a random point (Figure 4d).

The results of our experiment are presented in Figures 4 (a, b, c, d):

a – unpainted drawing, the blank for our experiment;

b – every area has its own unique color;

c – results of tearing up;

d – (optional) in each piece there is a distinct point that determines the color of the area.

The calculation results are listed in the Table 2.

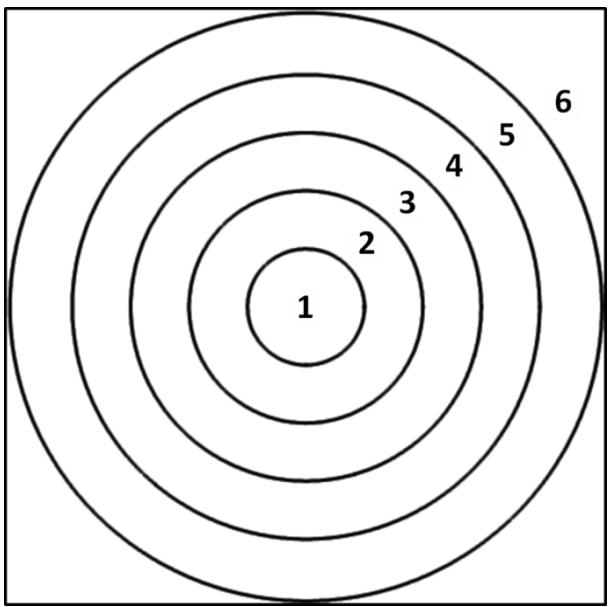

*a*

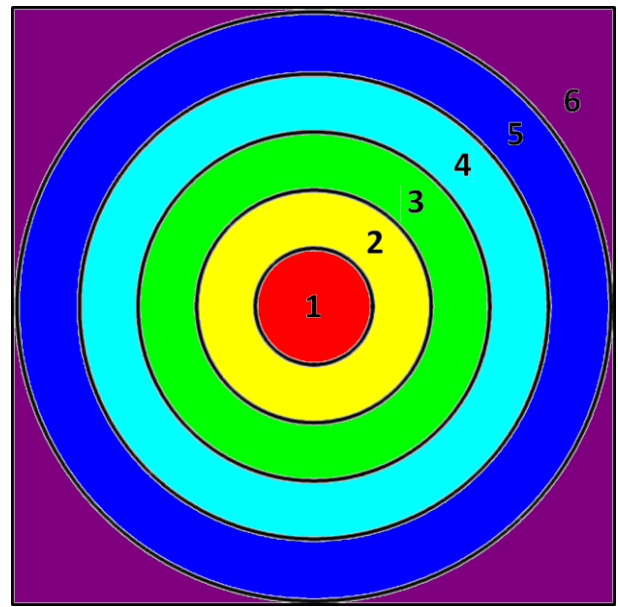

*b*

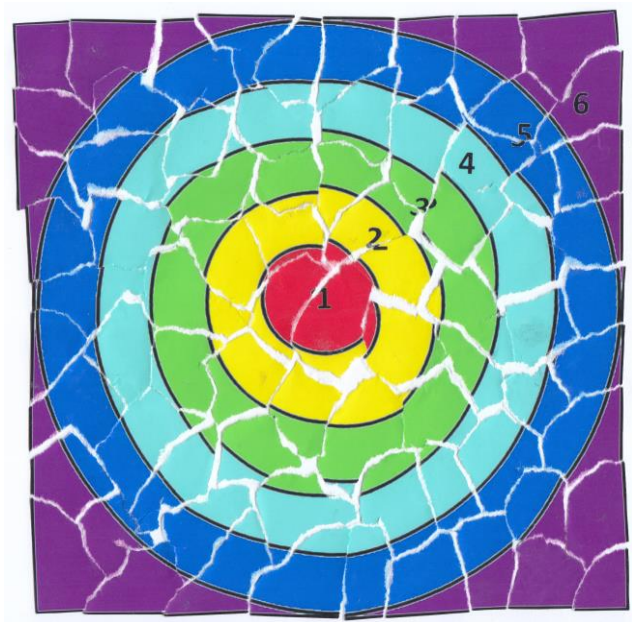

*c*

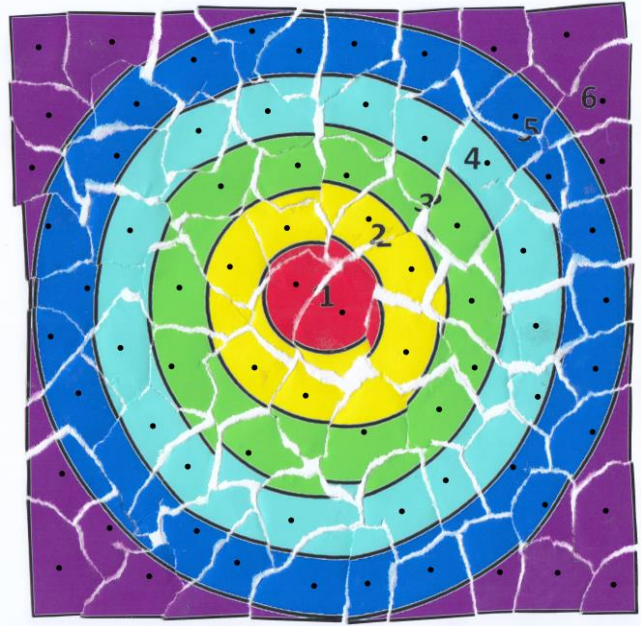

*d*

***Figure 4 (a, b, c, d)***

Table 2. The results of the second experiment

| Field Number k | Field Color | The Number of Results in the Field | Total Result (including field k) | $A_k = 4 \cdot \dfrac{N_k}{N_6}$ |
|---|---|---|---|---|
| 1 | red | 2 | 2 | 0.11 |
| 2 | yellow | 7 | 9 | 0.48 |
| 3 | green | 11 | 20 | 1.07 |
| 4 | cyan | 15 | 35 | 1.87 |
| 5 | blue | 24 | 59 | 3.15 |
| 6 | purple | 16 | 75 | 4.00 |

Here $N_6$ is the total number of points (pieces of paper). Take a look at the fifth line: $\pi$ appears here again. The last cell of the sixth line contains the area of the square in units of R. Here R is the radius of the largest (inscribed in the square) circle, or the half-side of the square: $d = 2 \cdot R$.

On the Figure 5a and 5b you can see the plot curve of the area of a circle on the radius $A = f(R)$ and linearized dependence $A = \varphi(R^2)$ (according to the Table 2). A straight line (passing through the origin) shows the degree of coincidence of the theory with data obtained by the Monte Carlo method.

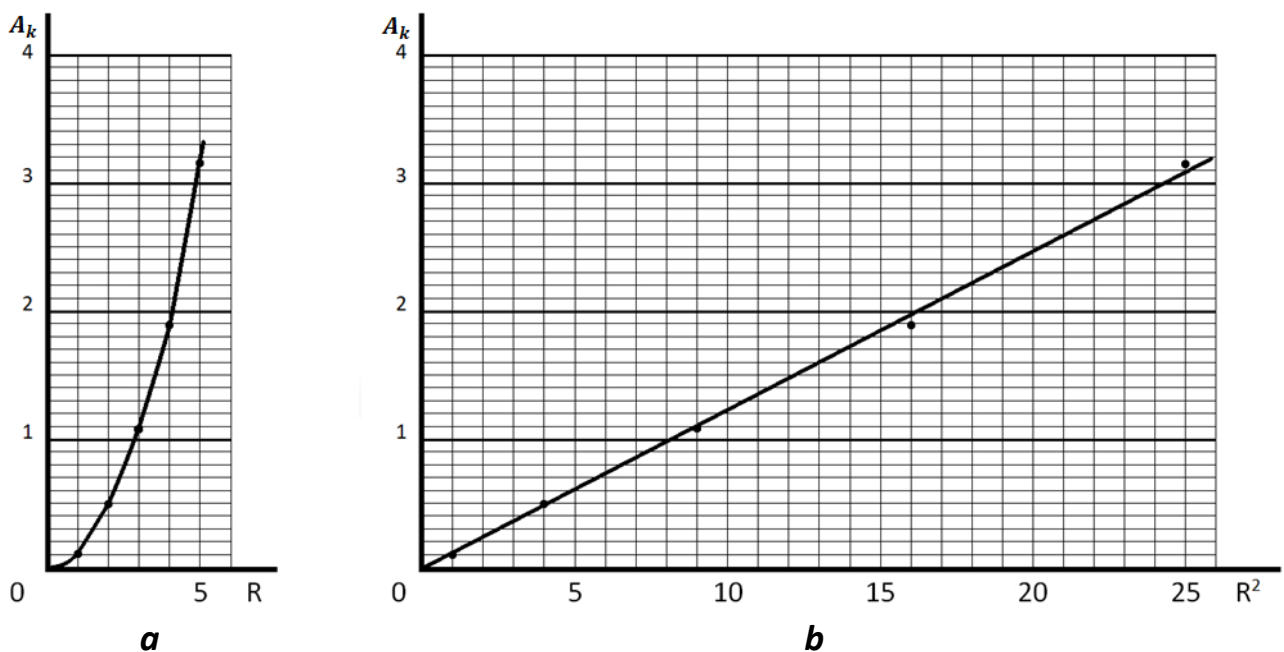

*Figure 5 (a, b). The results of the second experiment*
*a) The Dependence of the Circle Area A on the Radius R*
*b) The Dependence of the Circle Area A on the Radius Squared $R^2$*

## Conclusion

This version of the Monte Carlo method allows experimentally to prove the quadratic dependence of the area of the circle on the radius, and also to calculate $\pi$ as the coefficient of proportionality in the famous formula of the area of the circle.

Similarly, we can calculate the square roots, for example, a root of three, comparing the area of the square $A_1 = a^2$ and the equilateral triangle (Figure 6) with the same side $A_0 = a^2 \cdot \frac{\sqrt{3}}{4}$, tearing up this picture into small pieces and then counting the number of black and white ones:

$$\frac{A_0}{A_1} = \frac{a^2 \cdot \frac{\sqrt{3}}{4}}{a^2} = \frac{\sqrt{3}}{4}.$$

And as a result from here:

$$\sqrt{3} = 4 \cdot \frac{A_0}{A_1}.$$

It remains only to cut out a square from paper, to conduct an experiment and calculations (Appendix 2).

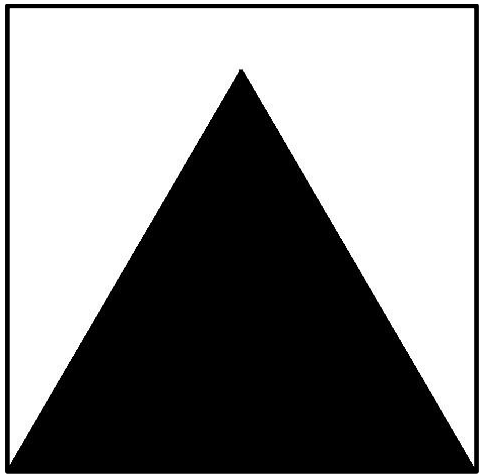

*Figure 6. The equilateral triangle inside the square*

The following problems await solutions by the Monte Carlo Method: the proof of formula for the area of a square; the dependence of the area of the triangle on its height, determining the relationship between the diameter (radius) and circumference of a circle.

I take this opportunity to offer teachers to find out various ways of including the Monte Carlo method in the school (university) physics course. Incorporation of Monte Carlo techniques into science and mathematics education is the old and still unresolved pedagogical problem [32].

One might say jokingly that bureaucrats of the world produce a huge amount of the paper documents. Finally we have found how to apply the mountains of this useless stuff for scientific and educational purposes.

**Appendix 1**

# MONTE CARLO METHOD

### Experiment 1. Calculation of Pi

| k | Total Number $N_0$ | Inside the Circle N | $\pi_k = 4 \cdot \dfrac{N}{N_0}$ | $|<\pi> - \pi_k|$ | $(<\pi> - \pi_k)^2$ |
|---|---|---|---|---|---|
| 1 | | | | | |
| 2 | | | | | |
| 3 | | | | | |
| 4 | | | | | |
| 5 | | | | | |
| | | | $<\pi>$= | | $\sum_k (<\pi> - \pi_k)^2 =$ |

Uncertainty $\Delta \pi = t_{st} \cdot \sqrt{\dfrac{\sum_k (<\pi> - \pi_k)^2}{N(N-1)}}$ =__________, ($t_{st}$ = _______ if $k_{max}$ = ___) and p = _____%)

**Finally π = _______________________________**

### Experiment 2. The Proof of $A = \pi r^2$

| Field Number k | Field Color | The Number of Results in the Field | Total Result (including field k) | $A_k = 4 \cdot \dfrac{N_k}{N_6}$ |
|---|---|---|---|---|
| 1 | ☐ | | | |
| 2 | ☐ | | | |
| 3 | ☐ | | | |
| 4 | ☐ | | | |
| 5 | ☐ | | | |
| 6 | ☐ | | | |

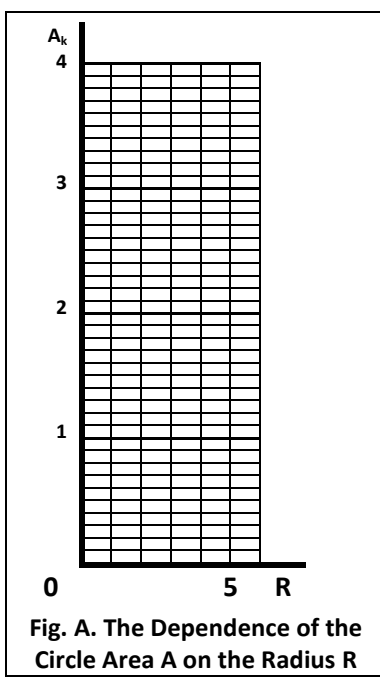

Fig. A. The Dependence of the Circle Area A on the Radius R

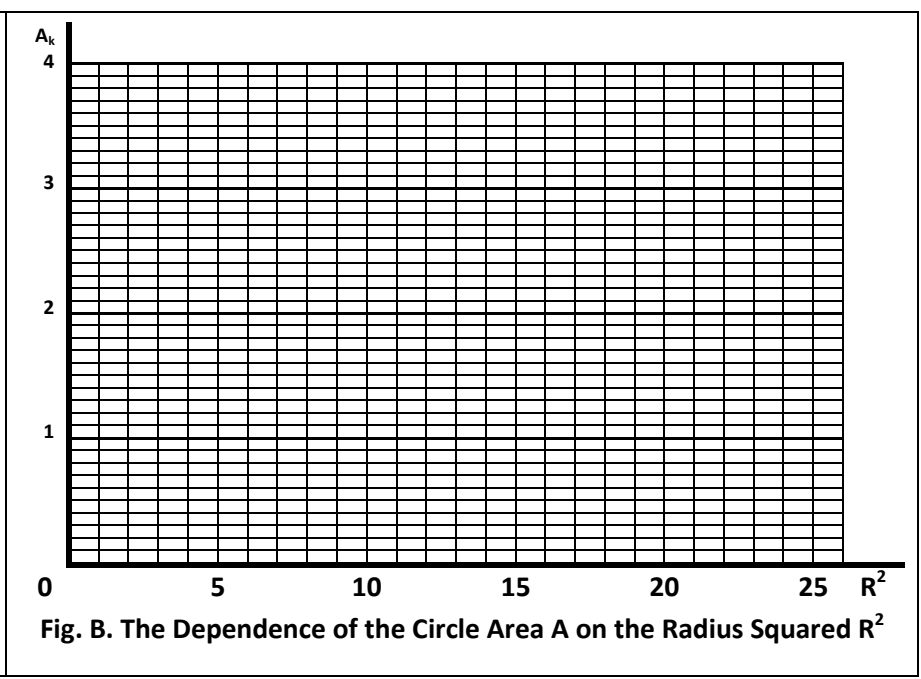

Fig. B. The Dependence of the Circle Area A on the Radius Squared $R^2$

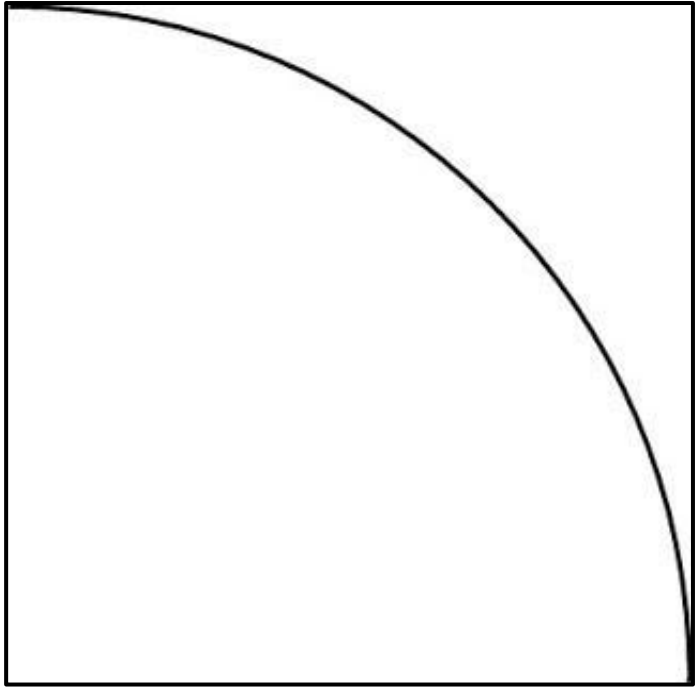
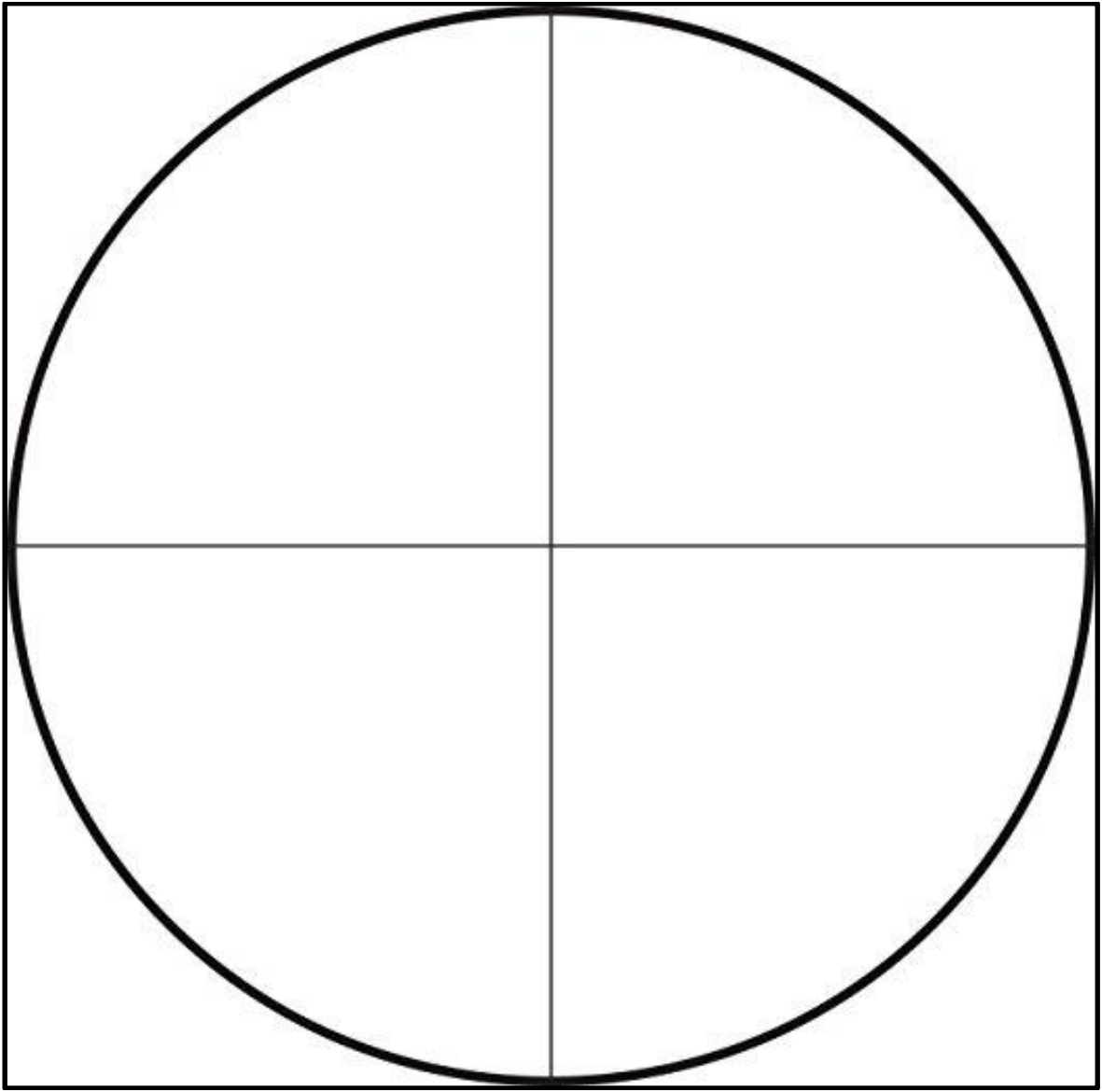

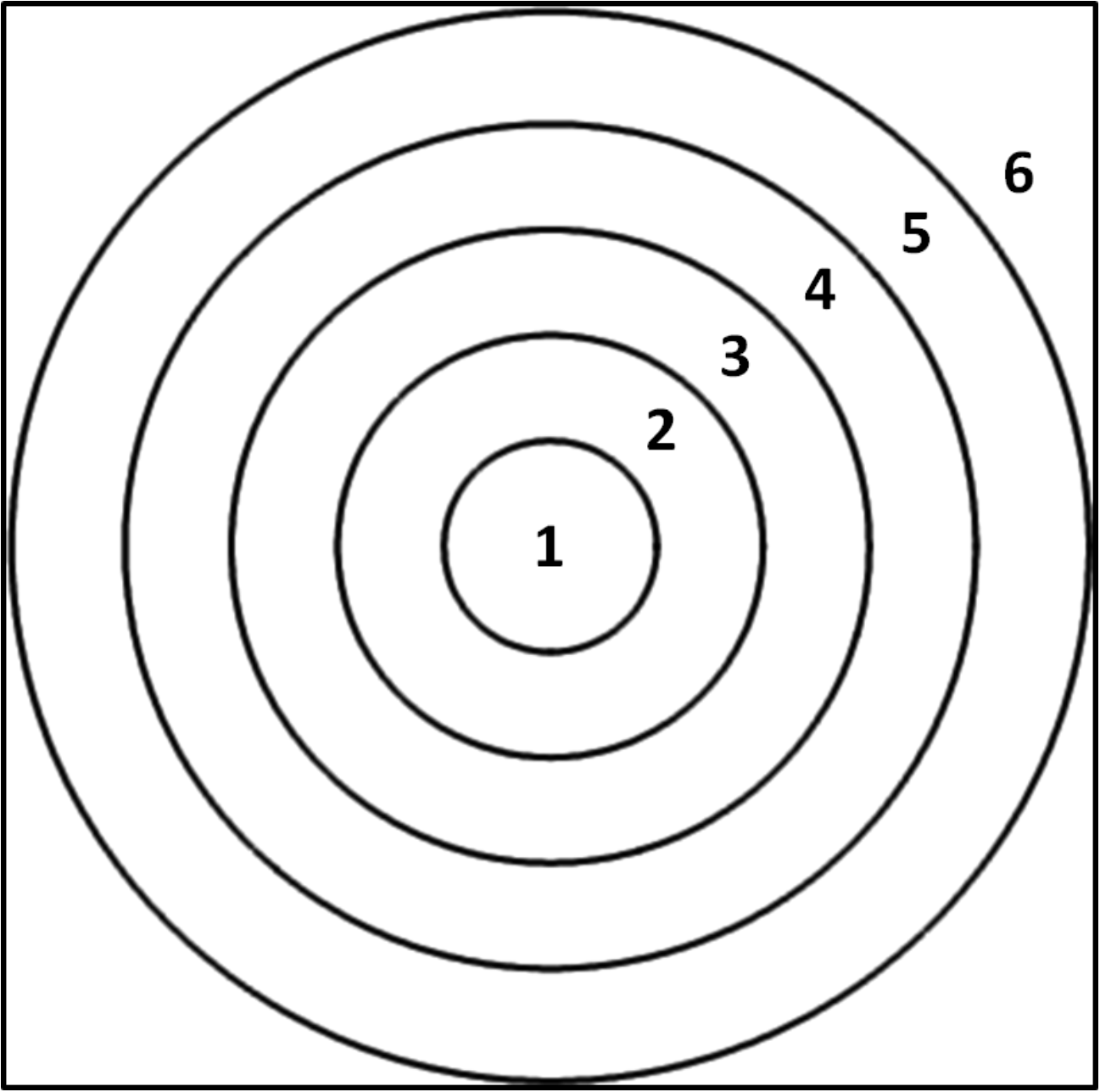

# MONTE CARLO METHOD ($\sqrt{3}$)

| k | Total Number $N_0$ | Inside the Triangle N | $x_k = 4 \cdot \dfrac{N}{N_0}$ | $|<x> - x_k|$ | $(<x> - x_k)^2$ |
|---|---|---|---|---|---|
| 1 | | | | | |
| 2 | | | | | |
| 3 | | | | | |
| 4 | | | | | |
| 5 | | | | | |
| | | | <x>= | | $\sum_k (<x> - x_k)^2 =$ |

Uncertainty $\Delta x = t_{st} \cdot \sqrt{\dfrac{\sum_k (<x> - x_k)^2}{N(N-1)}} =$ __________, ($t_{st} =$ __________ if $k_{max} =$ ____ and p = ____%)

Finally $\sqrt{3} =$ ______________________________________

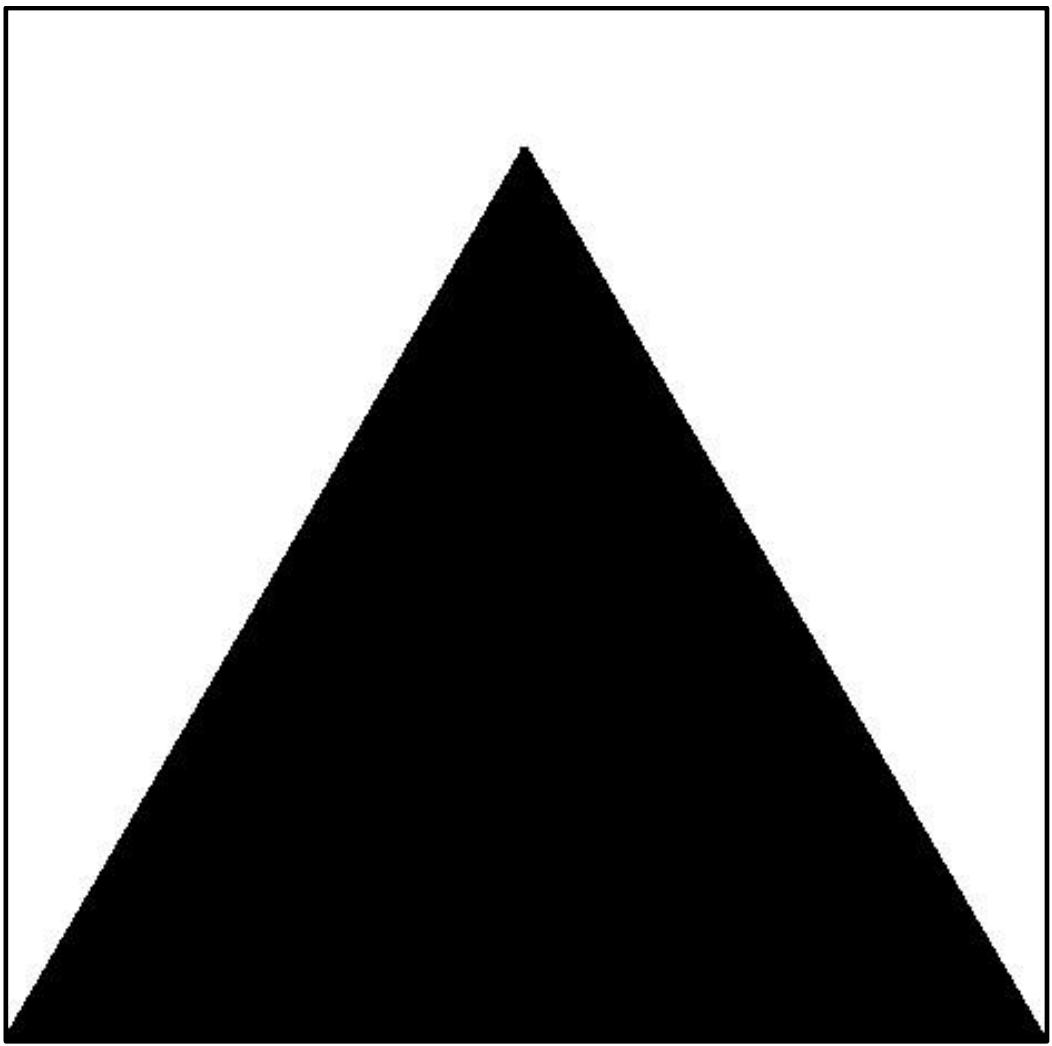